\documentclass[apjl]{emulateapj}
\usepackage{comment}
\usepackage{ifthen}

\newcommand{\ctbd}[1]{}

\newcommand{\lc}{light curve}
\newcommand{\lcs}{light curves}
\newcommand{\Lc}{Light curve}

\newcommand{\band}[1]{\ensuremath{#1}-band}

\newcommand{\kms}{\ensuremath{\rm km\,s^{-1}}}
\newcommand{\ms}{\ensuremath{\rm m\,s^{-1}}}

\newcommand{\gcmc}{\ensuremath{\rm g\,cm^{-3}}}
\newcommand{\ergscmsq}{\ensuremath{\rm erg\,s^{-1}\,cm^{-2}}}

\newcommand{\vsini}{\ensuremath{v \sin{i}}}
\newcommand{\feh}{\ensuremath{\rm [Fe/H]}}

\newcommand{\rsun}{\ensuremath{R_\sun}}
\newcommand{\msun}{\ensuremath{M_\sun}}
\newcommand{\lsun}{\ensuremath{L_\sun}}

\newcommand{\rstar}{\ensuremath{R_\star}}
\newcommand{\mstar}{\ensuremath{M_\star}}
\newcommand{\lstar}{\ensuremath{L_\star}}

\newcommand{\teffstar}{\ensuremath{T_{\rm eff\star}}}
\newcommand{\rhostar}{\ensuremath{\rho_\star}}
\newcommand{\loggstar}{\ensuremath{\log{g_{\star}}}}

\newcommand{\rpl}{\ensuremath{R_{p}}}
\newcommand{\mpl}{\ensuremath{M_{p}}}

\newcommand{\rhopl}{\ensuremath{\rho_{p}}}

\newcommand{\arstar}{\ensuremath{a/\rstar}}
\newcommand{\zrstar}{\ensuremath{\zeta/\rstar}}

\newcommand{\rjup}{\ensuremath{R_{\rm J}}}
\newcommand{\mjup}{\ensuremath{M_{\rm J}}}

\newcommand{\refsec}[1]{\mbox{\S\ \ref{sec:#1}}}

\newcommand{\reffigl}[1]{Figure~\ref{fig:#1}}
\newcommand{\refsecl}[1]{\mbox{Section \ref{sec:#1}}}

\newcommand{\reftabl}[1]{Table~\ref{tab:#1}}

\newcommand{\hatcurhtr}{HATS563-036}                                   
\newcommand{\hatcurCCra}{\ensuremath{11^{\mathrm h}42^{\mathrm m}06.12{\mathrm s}}}                                  
\newcommand{\hatcurCCdec}{\ensuremath{-23{\arcdeg}21{\arcmin}17.4{\arcsec}}}                                 
\newcommand{\hatcurCCtwomass}{2MASS~11420608-2321174}                  
\newcommand{\hatcurCCgsc}{GSC~6652-00186}                              
\newcommand{\hatcurCCapassmVshort}{\ensuremath{12.05}}                
\newcommand{\hatcurCCapassmV}{\ensuremath{12.053\pm0.017}}             
\newcommand{\hatcurCCapassmB}{\ensuremath{12.740\pm0.066}}              
\newcommand{\hatcurCCapassmg}{\ensuremath{12.356\pm0.052}}             
\newcommand{\hatcurCCapassmr}{\ensuremath{11.866\pm0.045}}             
\newcommand{\hatcurCCapassmi}{\ensuremath{11.748\pm0.042}}             
\newcommand{\hatcurCCtwomassJmag}{\ensuremath{10.924\pm0.024}}         
\newcommand{\hatcurCCtwomassHmag}{\ensuremath{10.634\pm0.022}}         
\newcommand{\hatcurCCtwomassKmag}{\ensuremath{10.580\pm0.023}}         
\newcommand{\hatcurLCrprstar}{\ensuremath{0.1288\pm0.0020}}            
\newcommand{\hatcurLCimp}{\ensuremath{0.633_{-0.023}^{+0.020}}}        
\newcommand{\hatcurLCzeta}{\ensuremath{23.98\pm0.16}}                  
\newcommand{\hatcurLCdur}{\ensuremath{0.1007\pm0.0012}}                
\newcommand{\hatcurLCingdur}{\ensuremath{0.0181\pm0.0011}}             
\newcommand{\hatcurLCP}{\ensuremath{3.446459\pm0.000004}}              
\newcommand{\hatcurLCPprec}{\ensuremath{3.4464588}}                    
\newcommand{\hatcurLCPshort}{\ensuremath{3.4465}}                      
\newcommand{\hatcurLCT}{\ensuremath{2455241.28722\pm0.00071}}          
\newcommand{\hatcurSPCiteff}{\ensuremath{5870\pm100}}                  
\newcommand{\hatcurSPCizfeh}{\ensuremath{-0.06\pm0.12}}                
\newcommand{\hatcurSPCizfehshort}{\ensuremath{-0.06}}                  
\newcommand{\hatcurSPCilogg}{\ensuremath{4.52\pm0.15}}                 
\newcommand{\hatcurSPCivsin}{\ensuremath{2.52\pm0.5}}                  
\newcommand{\hatcurSPCivmac}{\ensuremath{NULL}}                        
\newcommand{\hatcurSPCivmic}{\ensuremath{NULL}}                        
\newcommand{\hatcurSPCiiteff}{\ensuremath{5814\pm100}}                 
\newcommand{\hatcurSPCiizfeh}{\ensuremath{-0.1\pm0.1}}                 
\newcommand{\hatcurSPCiizfehshort}{\ensuremath{-0.1}}                  
\newcommand{\hatcurSPCiilogg}{\ensuremath{4.40\pm0.08}}                
\newcommand{\hatcurSPCiivsin}{\ensuremath{2.60\pm0.5}}                 
\newcommand{\hatcurSPCiivmac}{\ensuremath{NULL}}                       
\newcommand{\hatcurSPCiivmic}{\ensuremath{NULL}}                       
\newcommand{\hatcurLBii}{\ensuremath{0.2560}}                          
\newcommand{\hatcurLBiii}{\ensuremath{0.3331}}                         
\newcommand{\hatcurLBig}{\ensuremath{0.5230}}                          
\newcommand{\hatcurLBiig}{\ensuremath{0.2574}}                         
\newcommand{\hatcurLBir}{\ensuremath{0.3387}}                          
\newcommand{\hatcurLBiir}{\ensuremath{0.3343}}                         
\newcommand{\hatcurISOmshort}{\ensuremath{0.99}}                       
\newcommand{\hatcurISOmlong}{\ensuremath{0.986\pm0.054}}               
\newcommand{\hatcurISOrshort}{\ensuremath{1.04}}                       
\newcommand{\hatcurISOrlong}{\ensuremath{1.038_{-0.075}^{+0.128}}}     
\newcommand{\hatcurISOlogg}{\ensuremath{4.40\pm0.08}}                  
\newcommand{\hatcurISOlum}{\ensuremath{1.10_{-0.18}^{+0.33}}}          
\newcommand{\hatcurISOmv}{\ensuremath{4.73\pm0.24}}                    
\newcommand{\hatcurISOage}{\ensuremath{6.0\pm2.8}}                     
\newcommand{\hatcurISOMK}{\ensuremath{3.21\pm0.22}}                    
\newcommand{\hatcurISOspec}{G}                                         
\newcommand{\hatcurRVK}{\ensuremath{254.1\pm33.1}}                     
\newcommand{\hatcurRVk}{\ensuremath{-0.064_{-0.064}^{+0.041}}}         
\newcommand{\hatcurRVh}{\ensuremath{0.097_{-0.073}^{+0.102}}}          
\newcommand{\hatcurRVjitterB}{\ensuremath{61.5}}                       
\newcommand{\hatcurRVeccen}{\ensuremath{0.120\pm0.092}}                
\newcommand{\hatcurPPi}{\ensuremath{85.6_{-1.4}^{+0.6}}}               
\newcommand{\hatcurPPlogg}{\ensuremath{3.43\pm0.07}}                   
\newcommand{\hatcurPPar}{\ensuremath{9.20\pm0.82}}                     
\newcommand{\hatcurPParel}{\ensuremath{0.0444\pm0.0008}}               
\newcommand{\hatcurPPrho}{\ensuremath{1.03\pm0.25}}                    
\newcommand{\hatcurPPmshort}{\ensuremath{1.86}}                        
\newcommand{\hatcurPPmlong}{\ensuremath{1.855_{-0.196}^{+0.262}}}      
\newcommand{\hatcurPPrshort}{\ensuremath{1.30}}                        
\newcommand{\hatcurPPrlong}{\ensuremath{1.302_{-0.098}^{+0.162}}}      
\newcommand{\hatcurPPmrcorr}{\ensuremath{0.53}}                        
\newcommand{\hatcurPPteff}{\ensuremath{1359_{-59}^{+89}}}              
\newcommand{\hatcurPPtheta}{\ensuremath{0.127\pm0.014}}                
\newcommand{\hatcurPPfluxavg}{\ensuremath{7.70_{-1.22}^{+2.42}}}       
\newcommand{\hatcurPPfluxavgdim}{\ensuremath{8}}                       
\newcommand{\hatcurXdist}{\ensuremath{303_{-23}^{+38}}}                

\newcommand{\hatcur}{HATS-1}
\newcommand{\hatcurb}{HATS-1b}

\newcommand{\hatcurSPCversion}{i}                                       
\newcommand{\hatcurSPCteff}{\ifthenelse{\equal{\hatcurSPCversion}{i}}{\hatcurSPCiteff}{\hatcurSPCiiteff}}
\newcommand{\hatcurSPCzfeh}{\ifthenelse{\equal{\hatcurSPCversion}{i}}{\hatcurSPCizfeh}{\hatcurSPCiizfeh}}
\newcommand{\hatcurSPCzfehshort}{\ifthenelse{\equal{\hatcurSPCversion}{i}}{\hatcurSPCizfehshort}{\hatcurSPCiizfehshort}}
\newcommand{\hatcurSPClogg}{\ifthenelse{\equal{\hatcurSPCversion}{i}}{\hatcurSPCilogg}{\hatcurSPCiilogg}}
\newcommand{\hatcurSPCvsin}{\ifthenelse{\equal{\hatcurSPCversion}{i}}{\hatcurSPCivsin}{\hatcurSPCiivsin}}
\newcommand{\hatcurSPCvmac}{\ifthenelse{\equal{\hatcurSPCversion}{i}}{\hatcurSPCivmac}{\hatcurSPCiivmac}}
\newcommand{\hatcurSPCvmic}{\ifthenelse{\equal{\hatcurSPCversion}{i}}{\hatcurSPCivmic}{\hatcurSPCiivmic}}

\newcommand{\hatcurisoshort}{YY}
\newcommand{\hatcurisocite}{yi:2001}
\newcommand{\hatcurlumind}{\arstar}
\newcommand{\hatcurjhkfilset}{ESO}

\usepackage{lineno}
\usepackage{lipsum}

\newboolean{emulateapj}
\setboolean{emulateapj}{true}

\newboolean{rvtablelong}
\setboolean{rvtablelong}{true}

\newboolean{astroph}
\setboolean{astroph}{true}

\shortauthors{Penev et al.}
\shorttitle{\hatcur\lowercase{b}}

\ifthenelse{\boolean{emulateapj}}{
    
}{
    
}

\ifthenelse{\boolean{emulateapj}}{
    \newcommand{\titlestar}{$\star$}
}{
    \newcommand{\titlestar}{\star}
}

\begin{document}

\title{\hatcur\lowercase{b}: The First Transiting Planet Discovered by the
	HATSouth Survey \altaffilmark{$\dagger$}}
\author{
K.~Penev\altaffilmark{1,2},
G.~\'A.~Bakos\altaffilmark{1,2,\titlestar},
D.~Bayliss\altaffilmark{3},
A.~Jord\'an\altaffilmark{4},
M.~Mohler\altaffilmark{5},
G.~Zhou\altaffilmark{3},
V.~Suc\altaffilmark{4},
M.~Rabus\altaffilmark{4},
J.~D.~Hartman\altaffilmark{1,2},
L.~Mancini\altaffilmark{5},
B.~B\'eky\altaffilmark{2},
Z.~Csubry\altaffilmark{1,2},
L.~Buchhave\altaffilmark{6},
T.~Henning\altaffilmark{5},
N.~Nikolov\altaffilmark{5},
B.~Cs\'ak\altaffilmark{5},
R.~Brahm\altaffilmark{4},
N.~Espinoza\altaffilmark{4},
P.~Conroy\altaffilmark{3},
R.~W.~Noyes\altaffilmark{2},
D.~D.~Sasselov\altaffilmark{2},
B.~Schmidt\altaffilmark{3},
D.~J.~Wright,\altaffilmark{7},
C.~G.~Tinney,\altaffilmark{7},
B.~C.~Addison,\altaffilmark{7},
J.~L\'az\'ar\altaffilmark{8},
I.~Papp\altaffilmark{8},
P.~S\'ari\altaffilmark{8},
}
\altaffiltext{1}{Department of Astrophysical Sciences,
	Princeton University, NJ 08544, USA;
	email: kpenev@astro.princeton.edu}

\altaffiltext{2}{Harvard-Smithsonian Center for Astrophysics,
	Cambridge, MA, USA}

\altaffiltext{$\star$}{Alfred P.~Sloan Research Fellow}

\altaffiltext{3}{The Australian National University, Canberra,
	Australia}

\altaffiltext{4}{Departamento de Astronom\'ia y Astrof\'isica, Pontificia
	Universidad Cat\'olica de Chile, Av.\ Vicu\~na Mackenna 4860, 7820436 Macul,
	Santiago, Chile}

\altaffiltext{5}{Max Planck Institute for Astronomy, Heidelberg,
	Germany}

\altaffiltext{6}{Niels Bohr Institute, Copenhagen University, Denmark}

\altaffiltext{7}{Exoplanetary Science Group, School of Physics, University of
	New South Wales. 2052. Australia}

\altaffiltext{8}{Hungarian Astronomical Association, Budapest,
	Hungary}

\altaffiltext{$\dagger$}{
The HATSouth network is operated by a
collaboration consisting of Princeton University (PU), the Max Planck
Institute f\"ur Astronomie (MPIA), and the Australian National
University (ANU).  The station at Las Campanas Observatory (LCO) of the
Carnegie Institute, is operated by PU in conjunction with collaborators
at the Pontificia Universidad Cat\'olica de Chile (PUC), the station at
the High Energy Spectroscopic Survey (HESS) site is operated in
conjunction with MPIA, and the station at Siding Spring Observatory
(SSO) is operated jointly with ANU.
Based in part on observations made with the Nordic Optical Telescope,
operated on the island of La Palma in the Spanish Observatorio del
Roque de los Muchachos of the Instituto de Astrofisica de Canarias.
Based on observations made with the MPG/ESO 2.2\,m Telescope at the ESO
Observatory in La Silla. FEROS ID programmes: P087.A-9014(A),
P088.A-9008(A), P089.A-9008(A), P087.C-0508(A). GROND ID programme:
089.A-9006(A). This paper uses observations obtained with facilities
of the Las Cumbres Observatory Global Telescope.
}

\begin{abstract}

\setcounter{footnote}{10}

We report the discovery of \hatcurb{}, a transiting extrasolar planet
orbiting the moderately bright V=\hatcurCCapassmVshort\ \hatcurISOspec\ dwarf
star \hatcurCCgsc, and the first planet discovered by HATSouth, a global
network of autonomous wide-field telescopes. \hatcurb{} has a period of
$P\approx\hatcurLCPshort$\,d, mass of $\mpl \approx \hatcurPPmshort$\,\mjup,
and radius of $\rpl \approx \hatcurPPrshort$\,\rjup.  The host star has a
mass of \hatcurISOmshort\,\msun, and radius of \hatcurISOrshort\,\rsun.  The
discovery light curve of \hatcurb{} has near continuous coverage over several
multi-day periods, demonstrating the power of using a global network of
telescopes to discover transiting planets.
\setcounter{footnote}{0} \end{abstract}



\section{Introduction}
\label{sec:introduction}

The detection and study of extrasolar planets has become one of the fastest
developing fields in astrophysics. This has been fueled by the
extraordinarily rapid rate of exoplanetary discovery. The two most productive
discovery techniques have been the radial velocity (RV) method (measuring the
reflex RV signature of a planetary orbit in its parent star) and the transit
method (measuring the slight decrease in stellar brightness that occurs if
the planet happens to pass across its parent star's disk as viewed from the
Earth).

Even though among confirmed planets, much more were detected by the RV method
than the transit method, the latter enables many studies that are not
possible for RV detected planets, since for each transiting system a lot more
information is available than for an RV only system, like the planet radius,
an unambiguous mass and the inclination of the orbit relative to the line of
sight.

In addition, observations and measurements are possible for transiting
planets that cannot be done with RV-only planets. It is possible to measure
the misalignment between stellar spin and planetary orbits through the
Rossiter-McLaughlin (RM) effect \citep[e.g.,][and many others]{queloz:2000,
winn:2005, winn:2006, winn:2007, winn:2008, winn:2009, winn:2010a,
winn:2010b, winn:2011, narita:2007, narita:2008, narita:2009a, narita:2009b,
narita:2010a, narita:2010b}. Information can be obtained about the planetary
atmosphere by measuring the dependence of the transit depth (planet radius)
on wavelength \citep[e.g.,][]{charbonneau:2002, redfield:2008, snellen:2008,
sing:2008, sing:2009, sing:2011a, sing:2011b}. The surface brightness and
atmospheric temperature of the planet can be determined from observations of
the secondary eclipse, when the planet goes behind the star, and from the out
of eclipse variations in the combined planet plus star brightness
\citep[e.g.,][]{snellen:2007, snellen:2010, deming:2007, deming:2011,
charbonneau:2008, demory:2012}, as well as a wide array of other theoretical
and observational studies.

For these reasons, increasing the sample of known transiting extrasolar
planets, and extending the range of systems for which transits are detectable
is very valuable. The HATSouth network, described in detail in
\citet{bakos:2012:hs} (to be submitted), is designed to achieve both of these
goals. With telescopes situated at three sites in the southern hemisphere,
roughly $120^\circ$ apart in longitude, HATSouth is capable of detecting
transits at longer periods than any other ground based network presently in
operation. In addition, as detailed in \citet{bakos:2012:hs}, it is one of
the most sensitive wide--field, ground based searches to small planets. Since
a large part of the sky is being surveyed, many bright stars will be checked
for transits. As a consequence, the HATSouth survey is expected to find many
of its planets around bright stars, which makes the detailed follow up
studies mentioned above feasible.

In this paper we present the first planet to come from the HATSouth
survey: \hatcurb{}. Since this is the first transiting planet discovered by
th esurvey, we describe in detail the data
analysis methods and the procedures used to confirm the planetary nature of
the object and derive the stellar and planetary parameters.

The layout of the paper is as follows. In \refsecl{obs} we report the
detection of the photometric signal and the follow-up spectroscopic and
photometric observations of the host star, \hatcur{}.  In \refsecl{analysis}
we describe the analysis of the data, beginning with the determination of the
stellar parameters, continuing with a discussion of the methods used to rule
out nonplanetary, false positive scenarios which could mimic the photometric
and spectroscopic observations, and finishing with a description of our
global modeling of the photometry and radial velocities.  Our findings are
discussed in \refsecl{discussion}.

\section{Observations}
\label{sec:obs}

\subsection{Photometric detection}
\label{sec:detection}

\begin{figure}[!ht]
\plotone{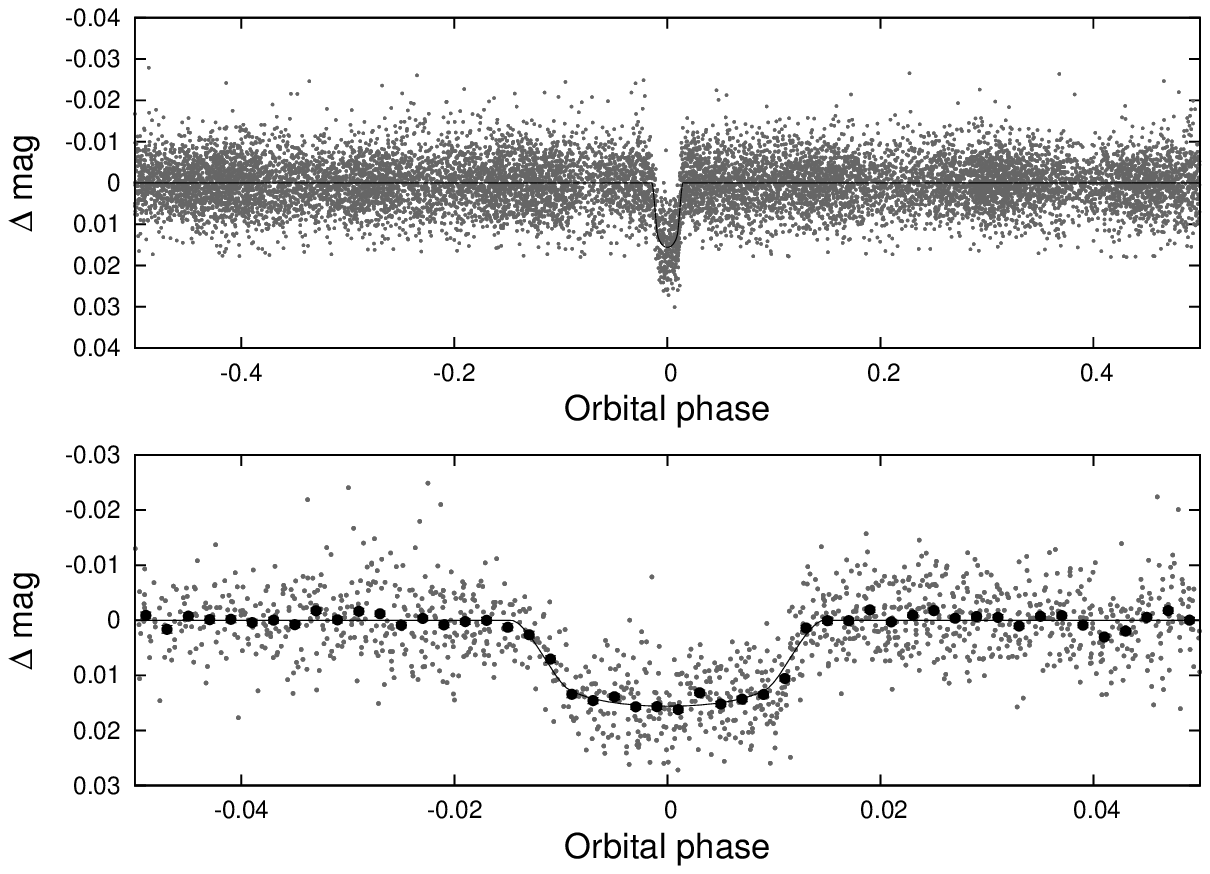}
\caption{
	Unbinned instrumental \band{r} \lc{} of \hatcur{} folded with
        the period $P = \hatcurLCPprec$\,days resulting from the
        global fit described in \refsecl{analysis}.  The solid line
        shows the best-fit transit model (see \refsecl{globmod}). In
        the lower panel we zoom-in on the transit; the dark filled
        points here show the light curve binned in phase using a
        bin-size of 0.002.
\label{fig:hatnet}}
\end{figure}

\ifthenelse{\boolean{emulateapj}}{
    \begin{deluxetable*}{llrrr}
}{
    \begin{deluxetable}{llrrr}
}
\tablewidth{0pc}
\tabletypesize{\scriptsize}
\tablecaption{
    Summary of photometric observations
    \label{tab:photobs}
}
\tablehead{
    \multicolumn{1}{c}{Facility}          &
    \multicolumn{1}{c}{Date(s)}             &
    \multicolumn{1}{c}{Number of Images}\tablenotemark{a}      &
    \multicolumn{1}{c}{Cadence (s)}\tablenotemark{b}         &
    \multicolumn{1}{c}{Filter}            \\
    &
    &
    &
    &
}
\startdata
HS-1 & 2010 Jan--2010 Aug & 6018 & 276 & Sloan~$r$ \\
HS-3 & 2010 Jan--2010 Aug & 5391 & 281 & Sloan~$r$ \\
HS-5 & 2010 Jan--2010 Aug &  655 & 271 & Sloan~$r$ \\
Swope/SITe3 & 2011 May 25     &   57 & 196 & Sloan~$i$ \\ 
FTS/Spectral & 2011 May 28    &   68 &  60 & Sloan~$r$ \\
FTS/Spectral & 2011 Jul 05    &   74 &  64 & Sloan~$i$ \\
MPG/ESO2.2/GROND & 2012 Jan 21    &   78 & 148 & Sloan~$g$ \\
MPG/ESO2.2/GROND & 2012 Jan 21    &   80 & 148 & Sloan~$r$ \\
MPG/ESO2.2/GROND & 2012 Jan 21    &   78 & 148 & Sloan~$i$ \\
[-1.5ex]
\enddata 
\tablenotetext{a}{
  Excludes images which were rejected as significant outliers in the
  fitting procedure.
}
\tablenotetext{b}{
  The mode time difference between consecutive points in each light
  curve. Due to visibility, weather, pauses for focusing, etc., none
  of the light curves have perfectly uniform time sampling.  
}
\ifthenelse{\boolean{emulateapj}}{
    \end{deluxetable*}
}{
    \end{deluxetable}
}

\reftabl{photobs} summarizes the HATSouth discovery observations of
\hatcurb{}. Observations were obtained over an eight month period from 2010
Jan through 2010 Aug using the HS1 (HATSouth 1) instrument in Chile, the HS3
instrument in Namibia, and the HS5 instrument in Australia. The observations,
reduction, and analysis of the HATSouth data were carried out as described in
\citet{bakos:2012:hs}. We detected a significant transit signal in the \lc{}
of \hatcurCCgsc{} (also known as \hatcurCCtwomass{}; $\alpha = \hatcurCCra$,
$\delta = \hatcurCCdec$; J2000; V=\hatcurCCapassmV\,mag APASS DR5,
e.g.~\citealp{henden:2009}; see \reffigl{hatnet}).

\subsection{Spectroscopy}
\label{sec:spec}

\reftabl{specobssummary} summarizes the follow-up spectroscopic observations
which we obtained for \hatcur{}. Initial medium and low-resolution
``reconnaissance'' observations were obtained using the Wide Field
Spectrograph \citep[WiFeS;][]{dopita:2007} on the ANU~2.3\,m telescope. These
observations were used to rule out various common false positive scenarios
(blends between an eclipsing binary and a giant star, or F-M type binary
systems). We then obtained higher resolution, and higher RV precision
observations with the Coralie spectrograph on the Euler 1.2\,m telescope at
La Silla Observatory, FEROS on the MPG/ESO 2.2\,m telescope at La Silla,
FIES \citep{djupvik:2010} on the 2.5\,m Nordic Optical Telescope, and CYCLOPS
on the 3.9\,m Anglo-Australian Telescope (AAT) at Siding Spring Observatory
(SSO). Below we describe the observations and reduction techniques for each
spectroscopic instrument used.

%
\subsubsection{ANU~2.3\,m/WiFeS}\label{sec:wifesobs}

High signal to noise (S/N), medium resolution reconnaissance spectroscopy was
performed with WiFeS on the ANU 2.3\,m telescope. WiFeS is an image slicing
integral field spectrograph with 25 slitlets each sampled at $0.5^{\prime\prime}$.
Our observations, made with the red arm of the spectrograph, employed the
R7000 grating and the RT480 dichroic, giving a spectral resolution of
$\lambda / \Delta \lambda = 7\,000$ and velocity dispersion of $21.6\,\kms\,
\text{pixel}^{-1}$ in the wavelength region $5\,200 - 7\,000 \, \text{\AA}$. The
$4\text{K} \times 4\text{K}$ CCD was read out in $2 \times$ binning in the
spatial direction, with a detector gain of $0.9\,e^{-}\,\text{ADU}^{-1}$, 
and read noise of $5 \, e^{-} \, \text{s}^{-1} \, \text{pixel}^{-1}$.
Only half the CCD containing the stellar signal was read out to reduce the
readout time. RV standard star exposures were taken on each night during
twilight.

Reduction of the spectra was performed using the IRAF package CCDPROC, and
extracted in TWODSPEC\@. Spectra from the three brightest image slices were
extracted individually. Wavelength calibrations were performed with
bracketing Ne-Ar arc lamp exposures. Atmospheric Oxygen B band lines in the
region $6\,882-6\,906\, \text{\AA}$ were used to provide a first order wavelength
correction. Object spectra were cross correlated against RV standard star
spectra in the wavelength region $5\,270-5\,650 \, \text{\AA}$. Velocity
measurements were obtained for each permutation of image slice and standard
star spectra cross correlation, performed using the task FXCOR\@. The template
and object spectra were Fourier filtered to remove low and high frequency
noise, and a Gaussian fit was performed on the peak of the cross correlation
function to derive the velocity shift. The final RV measurements were
calculated as the mean of these velocities weighted according to the cross
correlation function peak and object spectrum S/N.

The RV root mean square (RMS) scatter of multi-epoch WiFeS observations is
$1\,\kms$, as measured from RV standard stars. We scheduled our observations
at orbital phase quadrature of the candidate, where the velocity difference
is the greatest. The results from five observations (with mean exposure time
of 400\,s) show no RV variation greater than $2\,\kms$, indicating that the
candidate was not an unblended eclipsing binary.

We also obtained a single (700\,s exposure) low resolution ($\lambda / \Delta
\lambda = 3\,000$) spectrum which was used to estimate $\teffstar$ and
$\loggstar$.  We use the blue arm of WiFeS with the B3000 grating and RT560
dichroic, which gives a wavelength coverage of $3\,500-6\,000\, \text{\AA}$
at $\lambda / \Delta \lambda = 3\,000$ resolution.  Spectrophotometric
standard stars from \citet{hamuy:1994} are observed for flux calibration, and
a Ne-Ar lamp is used for wavelength calibration.  Reductions are performed
using the IRAF packages CCDPROC, KPNOSLIT and ONEDSPEC\@. The four brightest
image slices are summed, and the object spectrum is divided by a black body
white dwarf spectrum to remove the instrument sensitivity function.  Flux
calibration is performed according to the methodology set out in
\citet{bessell:1999}.  The spectrum is matched to a grid of synthetic MARCS
model atmospheres \citep{gustafsson:2008}, fitting for $\teffstar$,
$\loggstar$, [Fe/H], and interstellar reddening ($E(B-V)$), at 250\,K, 0.5\,dex,
and 0.01 magnitude intervals respectively.  Regions sensitive to changes
in $\log g$, such as the Balmer jump, the MgH feature at $4\,800 \,
\text{\AA}$, and the Mg b triplet at $5\,170\,\text{\AA}$, are weighted
preferentially. Based on the approximate $\loggstar$ and $\teffstar$, we find
that HATS-1 is a dwarf.

%
\subsubsection{Euler~1.2\,m/Coralie}\label{sec:coralieobs}

We used the CORALIE spectrograph mounted on the Swiss 1.2\,m Leonard Euler
telescope in the ESO La Silla Observatory to measure 6 radial velocities
(RVs) for \hatcurb{} during a run on 2011 May 19-20. CORALIE is a fiber-fed
Echelle spectrograph with a design similar to ELODIE \citep{baranne:1996} and
originally installed on April 1998 \citep{queloz:2001}. It was refurbished in
2007 to increase its optical efficiency; the refurbished instrument provides
a resolution of $\sim60\,000$. Spectra of our targets are taken in
simultaneous Thorium-Argon (ThAr) mode, whereby one of the fibers observes
the target and the other a ThAr lamp (hereafter the object and comparison
fibers, respectively).

An automated pipeline was developed at Pontificia Universidad Cat\'olica de
Chile to extract and calibrate the CORALIE spectra, following in large part
the reduction scheme for ELODIE described in \citet{baranne:1996}.  We
provide a brief description of the reduction procedure here, full details
will be given elsewhere. A night of observations include the following sets
of images: (i) science frames, where the object fiber observes the target and
the comparison fiber is illuminated by the ThAr lamp; (ii) ThAr frames, where
both fibers are illuminated by the ThAr lamp; (iii) flats, where each of the
fibers in turn is illuminated by a lamp at the beginning of the night. The
pipeline trims and bias subtracts all frames.  Flats are then median combined
and used to find and trace each of the Echelle orders; the traces are fit by
polynomials of order 4 and stored for extraction of the ThAr and science
frames. We extract 70 orders from the object fiber and 51 from the
comparison one, the wavelength coverage of the object fiber is $\approx
$3\,850--6\,900\,\AA.  All ThAr frames are then extracted using the traces
obtained from the flats, ThAr lines are identified and then fit by Gaussians.
Only lines from the list of \citet{lovis:2007} are used. With the line
centroids and wavelengths given in \citet{lovis:2007} in hand, a global
(i.e., for all orders simultaneously) wavelength solution of the form
$\lambda = P(x,m)/m$ is then fit, for the science and comparison fibers
independently, where $x$ is pixel position of ThAr lines along the dispersion
direction, $m$ is the Echelle order number and $P(x,m)$ is a polynomial of
order 3 in $x$ and 5 in $m$.  When fitting the wavelength solution we reject
lines iteratively until the rms around the fit is acceptable.

The spectrum in each Echelle order of the object fiber in the science frames
is extracted using the optimal extraction algorithm of \citet{marsh:1989}.
The comparison ThAr spectra are extracted via a simple extraction and are
used to calculate the velocity shift $\delta v$ of the comparison fiber of
each science frame with respect to the comparison fiber of the ThAr frame
nearest in time. The wavelength calibration for the orders corresponding to
the object fiber in the science frames is done by applying the wavelength
solution of the object fiber of the ThAr frame nearest in time shifted by
$\delta v$. Finally, we apply a barycentric correction computed at the
flux-weighted mean of the exposure and calculated using the JPL Solar System
ephemeris. With the spectra calibrated and corrected to the barycenter of the
Solar System, a cross-correlation function is calculated using a binary mask
as described in \citet{baranne:1996}.  A spectral typing module within the
pipeline gives estimates of $T_{\rm eff}$, $\log(g)$, $v\sin (i)$ and [Fe/H]
for each target, and we use these to choose an appropriate cross-correlation
mask within a set of three choices: G2, K5, M2\footnote{The masks are the
same as those used with the HARPS spectrograph, with the ``transparent''
regions of the masks made wider to match the lower resolution of CORALIE with
respect to HARPS\@. Once a mask is chosen for a given target it is always
maintained for further observations of it.}. The resulting CCF is fit with a
Gaussian and the mean returned by the fit is the measured RV\@.  Following
\citet{queloz:1995} we estimate the uncertainty in the measured RVs via the
scaling $\sigma_{\rm RV} = a + [(1.6+0.2w)b/sn]$, where $sn$ is the
signal-to-noise ratio at 5\,130\,\AA\ , $w$ is the width of the CCF measured
in pixels, and $a\approx 6$\,\ms\ and $b \approx 100$\,\ms\ are constants
that are determined via simulations and depend slightly on spectral type. We
have monitored the RV standard HD~72673 for $\approx$ 2 years, obtaining 40
RV measurements with S/N$>$60 per pixel at 5\,130\,\AA. This star is known to
have a long-term RV rms of $\sim2$\,\ms\ from HIRES/Keck observations (Andrew
Howard, private communication). Our CORALIE RV measurements have an RMS of
8\,\ms, which showcases the long-term velocity precision we can currently
achieve with CORALIE for bright targets. The observations of HD~72673 show
that even at high S/N we do not reach the level implied by the constant $a$
in the expression for $\sigma_{\rm RV}$ but rather we are limited to
$\sigma_{\rm RV, min} = 8$\,\ms, which we adopt as our lower limit in the
velocity uncertainty estimates.

%
\subsubsection{NOT~2.5\,m/FIES}\label{sec:fiesobs}

We obtained two observations of \hatcur{} using the FIbre-fed
\'Echelle Spectrograph (FIES) on the 2.5\,m Nordic Optical Telescope
(NOT) located on the island of La Palma \citep{djupvik:2010}. The FIES
observations were reduced to spectra following the procedure of
\citet{buchhave:2010}. We used these spectra to measure the
atmospheric parameters of \hatcur{}, including the effective
temperature $\teffstar$, the metallicity $\feh$, the projected
rotational velocity $\vsini$, and the stellar surface gravity
$\loggstar$, by cross-correlating the observations against a finely
sampled grid of synthetic spectra based on \cite{kurucz:2005} model
atmospheres. The procedure, called Stellar Parameter Classification
(SPC), will be described in detail in a forthcoming paper (Buchhave et
al., in preparation, see also \citealp{bakos:2012:hat34to37} for
results for other stars based on this method).

%
\subsubsection{ESO~2.2\,m/FEROS}\label{sec:ferosobs}


\ifthenelse{\boolean{emulateapj}}{
    \begin{deluxetable*}{llrrrr}
}{
    \begin{deluxetable}{llrr}
}
\tablewidth{0pc}
\tabletypesize{\scriptsize}
\tablecaption{
    Summary of spectroscopic observations
    \label{tab:specobssummary}
}
\tablehead{
    \multicolumn{1}{c}{Telescope/Instrument} &
    \multicolumn{1}{c}{Date Range}          &
    \multicolumn{1}{c}{Number of Observations} &
    \multicolumn{1}{c}{Resolution}          \\
}
\startdata
ANU 2.3\,m/WiFeS & 2011 May 10---15 & 5 & 7\,000 \\
ANU 2.3\,m/WiFeS & 2011 June 05 & 1 & 3\,000 \\
AAT 3.9\,m/CYCLOPS & 2012 Jan 5---12 & 6 & 70\,000 \\
Euler 1.2\,m/Coralie & 2011 May 19---20 & 6 & 60\,000 \\
NOT 2.5\,m/FIES & 2011 June 15---18 & 2 & 67\,000 \\
MPG/ESO 2.2\,m/FEROS & 2011 June 08--- 2012 April 18 & 21 & 48\,000 \\
[-1.5ex]
\enddata 
\ifthenelse{\boolean{emulateapj}}{
    \end{deluxetable*}
}{
    \end{deluxetable}
}

\begin{figure} [ht]
\plotone{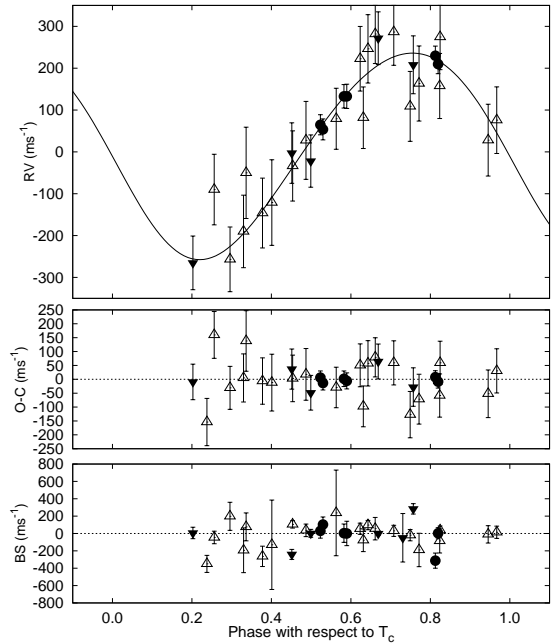}
\caption{
    {\em Top panel:} High-precision RV measurements for
    \hbox{\hatcur{}} from CORALIE (dark filled circles), FEROS
    (open triangles) and CYCLOPS (filled triangles) shown as a function of
    orbital phase, together with our best-fit model. Zero phase corresponds
    to the time of mid-transit. The center-of-mass velocity has been
    subtracted. {\em Second panel:} Velocity $O\!-\!C$ residuals from the
    best fit.  The error bars include a component from
    astrophysical/instrumental jitter allowed to differ for the three
    instruments. {\em Third panel:} Bisector spans (BS), with the mean value
    subtracted. Note the different vertical scales of the panels.
\label{fig:rvbis}}
\end{figure}

\ifthenelse{\boolean{emulateapj}}{
    \begin{deluxetable*}{lrrrrrr}
}{
    \begin{deluxetable}{lrrrrrr}
}
\tablewidth{0pc}
\tablecaption{
	Relative radial velocities and bisector span measurements of
	\hatcur{}.
	\label{tab:rvs}
}
\tablehead{
	\colhead{BJD} & 
	\colhead{RV\tablenotemark{a}} & 
	\colhead{\ensuremath{\sigma_{\rm RV}}\tablenotemark{b}} & 
	\colhead{BS} & 
	\colhead{\ensuremath{\sigma_{\rm BS}}} & 
        \colhead{Phase} &
        \colhead{Instrument}\\
	\colhead{\hbox{(2\,454\,000$+$)}} & 
	\colhead{(\ms)} & 
	\colhead{(\ms)} &
	\colhead{(\ms)} &
	\colhead{} &
        \colhead{} &
        \colhead{}
}
\startdata
$ 1701.46966 $ & $    64.78 $ & $    24.00 $ & $   30.4 $ & $   84.4 $ & $   0.523 $ & Coralie \\
$ 1701.49218 $ & $    53.78 $ & $    25.00 $ & $  103.3 $ & $   87.4 $ & $   0.530 $ & Coralie \\
$ 1701.67397 $ & $   132.78 $ & $    28.00 $ & $    3.7 $ & $  105.8 $ & $   0.583 $ & Coralie \\
$ 1701.69858 $ & $   132.78 $ & $    29.00 $ & $    0.2 $ & $  140.7 $ & $   0.590 $ & Coralie \\
$ 1702.46676 $ & $   229.78 $ & $    23.00 $ & $ -312.9 $ & $   87.2 $ & $   0.813 $ & Coralie \\
$ 1702.49046 $ & $   209.78 $ & $    23.00 $ & $    2.7 $ & $   64.3 $ & $   0.819 $ & Coralie \\
$ 1721.48200 $ & $  -190.09 $ & $    61.09 $ & $ -193.4 $ & $  257.4 $ & $   0.330 $ & FEROS \\
$ 1722.56100 $ & $   246.23 $ & $    53.72 $ & $   97.1 $ & $   53.8 $ & $   0.643 $ & FEROS \\
$ 1723.60400 $ & $    28.04 $ & $    59.75 $ & $  -10.1 $ & $  100.3 $ & $   0.946 $ & FEROS \\
$ 1724.61100 $ & $  -409.37 $ & $    58.84 $ & $ -350.1 $ & $   98.5 $ & $   0.238 $ & FEROS \\
$ 1726.63000 $ & $   157.47 $ & $    47.67 $ & $  -85.7 $ & $  139.2 $ & $   0.824 $ & FEROS \\
$ 1728.53900 $ & $  -146.13 $ & $    56.44 $ & $ -263.5 $ & $  117.4 $ & $   0.377 $ & FEROS \\
$ 1735.51300 $ & $  -121.06 $ & $    81.73 $ & $ -130.1 $ & $  514.8 $ & $   0.401 $ & FEROS \\
$ 1736.57100 $ & $   286.48 $ & $    50.45 $ & $   30.3 $ & $   63.6 $ & $   0.708 $ & FEROS \\
$ 1737.46400 $ & $    75.76 $ & $    50.47 $ & $   14.1 $ & $   70.5 $ & $   0.967 $ & FEROS \\
$ 1738.46000 $ & $   -90.05 $ & $    57.78 $ & $  -46.1 $ & $   72.1 $ & $   0.256 $ & FEROS \\
$ 1901.84000 $ & $   281.61 $ & $    34.46 $ & $   55.4 $ & $  129.7 $ & $   0.661 $ & FEROS \\
$ 1911.84100 $ & $    79.18 $ & $    39.24 $ & $  236.8 $ & $  493.2 $ & $   0.563 $ & FEROS \\
$ 1932.13480 $ & $    -2.80 $ & $    39.60 $ & $ -240.8 $ & $   57.8 $ & $   0.451 $ & CYCLOPS \\
$ 1933.09769 $\tablenotemark{d} & $  -104.80 $ & $   115.00 $ & $  -49.3 $ & $  279.6 $ & $   0.731 $ & CYCLOPS \\
$ 1933.18812 $ & $   208.10 $ & $    33.62 $ & $  283.5 $ & $   59.9 $ & $   0.757 $ & CYCLOPS \\
$ 1936.86700 $ & $   274.56 $ & $    48.03 $ & $   37.1 $ & $   49.0 $ & $   0.824 $ & FEROS \\
$ 1938.16979 $ & $  -265.20 $ & $    21.28 $ & $    6.0 $ & $   66.2 $ & $   0.202 $ & CYCLOPS \\
$ 1939.19224 $ & $   -22.00 $ & $    15.65 $ & $    1.6 $ & $   45.9 $ & $   0.499 $ & CYCLOPS \\
$ 1943.22390 $ & $   271.60 $ & $    17.65 $ & $   -0.9 $ & $   22.0 $ & $   0.669 $ & CYCLOPS \\
$ 1991.75100 $ & $   108.73 $ & $    56.55 $ & $  -19.4 $ & $   65.4 $ & $   0.749 $ & FEROS \\
$ 2021.75000 $ & $   -33.57 $ & $    57.03 $ & $  105.1 $ & $   42.7 $ & $   0.453 $ & FEROS \\
$ 2024.79300 $ & $   -50.14 $ & $    90.16 $ & $   76.1 $ & $  160.6 $ & $   0.336 $ & FEROS \\
$ 2025.78200 $ & $   222.49 $ & $    46.75 $ & $   54.7 $ & $   64.5 $ & $   0.623 $ & FEROS \\
$ 2029.73900 $ & $   163.37 $ & $    65.29 $ & $ -189.6 $ & $  192.5 $ & $   0.772 $ & FEROS \\
$ 2031.54600 $ & $  -256.83 $ & $    46.88 $ & $  197.9 $ & $  161.3 $ & $   0.296 $ & FEROS \\
$ 2032.70200 $ & $    81.66 $ & $    40.39 $ & $  -77.7 $ & $  130.8 $ & $   0.631 $ & FEROS \\
$ 2035.65100 $ & $    27.54 $ & $    69.87 $ & $   35.9 $ & $   72.2 $ & $   0.487 $ & FEROS \\

	[-1.5ex]
\enddata
\tablenotetext{a}{
        The zero-point of these velocities is arbitrary. An overall
        offset $\gamma_{\rm rel}$ fitted separately to the FIES,
        Keck and CYCLOPS velocities in \refsecl{globmod} has been subtracted.
}
\tablenotetext{b}{
	Internal errors excluding the component of
        astrophysical/instrumental jitter considered in
        \refsecl{globmod}.
}
\ifthenelse{\boolean{emulateapj}}{
    \end{deluxetable*}
}{
    \end{deluxetable}
}


FEROS is the ``Fiber-fed Extended Range Optical Spectrograph'' located at the
2.2\,m MPG/ESO telescope at the La Silla Observatory, Chile. The instrument
covers the whole wavelength range between 3\,500 and 9\,200\,\AA\ separated into 39
spectral orders at a resolving power $\lambda/\Delta\lambda$ of 48\,000
\citep{kaufer:1998}, making the analysis of the radial velocity and activity
indicators such as Ca II H and K simultaneously possible \citep{setiawan:2003}.

FEROS is equipped with a double fiber system, providing the opportunity to
observe in object-sky or in object-calibration mode. The observations for
HATS-1 were taken in the object-calibration mode to guarantee the highest
achievable precision with FEROS\@. In this mode, a ThAr+Ne lamp spectrum is
recorded parallel to the target exposure, in order to calibrate the wavelength
simultaneously. This method is modeled on the ELODIE spectrograph
\citep{baranne:1996}. 

\citet{kaufer:1998} report in the commissioning of FEROS at the 1.52\,m
telescope at La Silla observatory, Chile, a RV measurement error of
$\sim$23\,\ms. An improved data reduction method \citep{setiawan:2000},
reduced this error to $\sim$10\,m/s, enabling FEROS to be used for planet
searches.

Between June 2011 and April 2012, a total of 21 spectra of HATS-1 were taken
with FEROS during MPI guaranteed time (PI M.~Mohler) and Chilean time (PI
A.~Jord\'an).

The data reduction, including bias- and flatfield-correction, wavelength
calibration and barycentric motion correction, was done using the MIDAS data
reduction pipeline at the telescope. 

The radial velocity of each spectrum was calculated by cross-correlating the
object spectrum with a synthetic one, generated with the program SPECTRUM
\citep{gray:1994} using Kurucz models \citep{kurucz:1993}. In order to find
the stellar parameters to synthesize the fitting spectrum for
cross-correlation, the object spectra were analyzed using the program SME
(\textit{Spectroscopy Made Easy}, \cite{valenti:1996}). Since the instrument
efficiency of FEROS drops significantly below 4\,000\,\AA\ and above
$\sim$7\,900\,\AA, only the spectral range between these limits was used for
cross-correlation. Due to the fact that Balmer lines, areas of telluric lines
and strong emission features can falsify the result, these areas were
carefully excluded in the analysis. In total, 26 out of 39 orders were used.
The RV was determined by fitting a Gaussian function to the final
cross-correlation, with the center of the Gaussian fit representing the RV of
the star. 

%
\subsubsection{AAT~3.9\,m/CYCLOPS}\label{sec:uclesobs}

High-precision radial velocity observations were obtained with the CYCLOPS
fiber-system feeding the UCLES spectrograph on the 3.9\,m AAT at SSO,
Australia. UCLES is a cross-dispersed Echelle spectrograph located at the
coud\'{e} focus with a 79\,l\,mm$^{-1}$ grating.  CYCLOPS is a Cassegrain
fiber-based integral field unit used to feed UCLES.\@ It consists of a 15
element fiber bundle which re-formats a $\sim3^{\prime\prime}$ diameter
aperture into a pseudo--slit $0.63^{\prime\prime}$ wide and 15 elements long.
It delivers a spectral resolution of $\lambda / \Delta \lambda \sim 70\,000$
over 19 Echelle orders, with a total wavelength range of
$4\,540-7\,340\,\mathrm{\AA}$.  

The EEV2 $2\mathrm{K}\times 4\mathrm{K}$ CCD is readout with $\times 2$
binning in the spatial direction and the normal readout speed, delivering
$1.3 \, e^{-} \,\mathrm{ADU} ^{-1}$ gain and a read-noise of $3.19 \, e^{-}
\,\mathrm{pix}^{-1}$.  Exposure times vary between 1\,200\,s and 2\,400\,s
depending on the brightness of the target and the weather conditions.  We
take a ThAr exposure before and after each science exposure to monitor the
stability of the spectrograph.

Data reduction is performed using custom MATLAB routines, which trace each
fiber and optimally extract each spectral order.  Wavelength calibration is
performed by linear interpolation between the ThAr exposures on either side
of the exposure.  Using the IRAF task FXCOR we cross-correlate each of the
individual fiber spectra against corresponding fiber spectra of a very high
signal-to-noise radial velocity standard star exposure.  Final velocities are
determined as the weighted mean of all fibers over all orders with 3 sigma
clipping.  Typically 12 fibers over 17 orders are used, with the two reddest
orders too heavily contaminated with telluric lines to deliver precise RV
measurements.  RV uncertainties are calculated as the standard deviation of
the scatter of the fibers over all the orders used.

%
\subsection{Photometric follow-up observations} \label{sec:phot}

\begin{figure}[!ht] \plotone{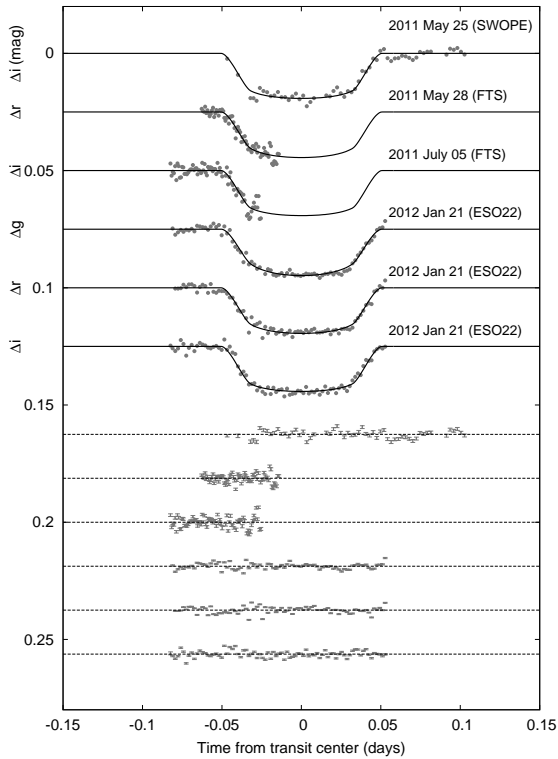} \caption{
	Unbinned instrumental Sloan $g$-, $r$-, and \band{i} transit \lcs{} of
	\hatcur{}. The dates and instruments used for each event are indicated. The
	light curves have been detrended using the EPD and TFA processes. Curves
	after the first are shifted for clarity. Our best fit is shown by the solid
	lines. Residuals from the fits are displayed at the bottom, in the same
	order as the top curves.
\label{fig:lc}} \end{figure}


Follow-up photometric observations of \hatcur{} were obtained using the SITe3
camera on the Swope 1.0\,m telescope at LCO, the Spectral instrument on the
2.0\,m Faulkes Telescope South (FTS) at SSO, and GROND on the MPG/ESO 2.2\,m
telescope at La Silla Observatory in Chile. \reftabl{photobs} summarizes these
observations. Below we describe the reduction procedures for each instrument
used.

%
\subsubsection{Swope~1\,m/SITe3}\label{sec:swope}

The SITe\#3 camera on the Swope~1\,m telescope is a 2048$\times$3150 detector
with a pixel scale of $0.435^{\prime\prime}$ and a field of view of
$14.8^\prime \times 22.8^\prime$. Observations of a field centered on
\hatcur{} were calibrated using standard methods and reduced to light curves
following a similar procedure to that used by \citet{bakos:2010:hat11} in the
analysis of photometric follow-up data for the HATNet project. Briefly, we
use a variation of the {\sc fitsh} package \citep{pal:2012:fitsh} to identify
sources on the images, cross-match them between images, perform aperture
photometry on the images using a range of apertures, collect light curves for
the detected sources, and perform an ensemble magnitude fit to correct the
light curves for variations that are common to multiple sources. Auxiliary
parameters describing the shape of the stellar profile, the position of each
source on each image, the sky background, and the local variation in the sky
background, are also recorded. These are used in the light curve fitting
procedure (\refsec{globmod}) to correct for systematic variations in the
target star that are not removed by the ensemble magnitude fit.

%
\subsubsection{FTS~2\,m/Spectral}\label{sec:FTS}

Two partial transits of \hatcurb{} were observed with the FTS, which is part
of the Las Cumbres Global Telescope (LCOGT) Network.  We used the
4K$\times$4K ``Spectral'' imaging camera, featuring $0.15^{\prime\prime}$
pixels and a $10^\prime$ FoV.  Exposures of 40\,s were taken in both \band{r}
(2011 May 28) and \band{i} (2011 July 5), binning $2\times2$ to reduce
readout time, and slightly de-focusing the telescope to avoid saturation.
Raw data is reduced via the automated reduction pipeline provided by LCOGT
for FTS data.  Photometry is performed by an automated pipeline based on
Source Extractor \citep{bertin:1996} aperture photometry and calibrated using
selected neighboring reference stars.  Details of these observations are set
out in Table \ref{tab:phfu}, and photometry of the two transit ingresses
observed is presented in Figure \ref{fig:hatnet}.

%
\subsubsection{ESO~2.2\,m/GROND}\label{sec:grond}

A complete transit of \hatcurb{} was observed on UT 2012 January 21 in the Sloan g
($\lambda = 4\,550$\,\AA), 
r ($\lambda = 6\,270$\,\AA) and i-band ($\lambda = 7\,630$\,\AA) of
the GROND instrument \citep{greiner:2008}, mounted on the MPG/ESO 2.2\,m
telescope at the ESO La Silla Observatory (Chile). The simultaneous multiband
time series was obtained applying a moderate telescope defocus to minimize
flat-fielding errors and to avoid saturation due to abrupt seeing decrease.
The field-of-view (FOV) of GROND in the three optical channels is $5.4^{\prime}
\times 5.4^{\prime}$ with a pixel scale of $0.158^{\prime\prime}$
$\rm{pixel^{-1}}$, which allowed monitoring of the target and a few nearby
reference stars. We acquired data for nearly 3.5\,hr time interval, covering
the predicted transit of \hatcurb.

The data reduction and analysis of the photometric data was performed using a
customized pipeline, following standard procedures. The masterbias and
masterflat frames were used to perform debias and flat-fielding of the science
images. DAOPHOT aperture photometry was performed as implemented in the IDL
APER function to obtain light curves of HATS-1 and the reference stars in
GROND's FOV with three aperture radii (27.5, 30 and 32.5 pixels).  Finally a
differential photometry was performed relative to the reference stars.

\ifthenelse{\boolean{emulateapj}}{ \begin{deluxetable*}{lrrrrr} }{
		\begin{deluxetable}{lrrrrr} } \tablewidth{0pc}
				\tablecaption{Differential photometry of
				\hatcur\label{tab:phfu}} \tablehead{ \colhead{BJD} &
				\colhead{Mag\tablenotemark{a}} &
				\colhead{\ensuremath{\sigma_{\rm Mag}}} &
				\colhead{Mag(orig)\tablenotemark{b}} & \colhead{Filter} &
				\colhead{Instrument} \\ \colhead{\hbox{~~~~(2\,400\,000$+$)~~~~}}
				& \colhead{} & \colhead{} & \colhead{} & \colhead{} &
				\colhead{} } \startdata $ 55287.81456 $ & $  -0.01671 $ & $   0.00299 $ & $ \cdots $ & $ r$ &         HS\\
$ 55308.49356 $ & $   0.00495 $ & $   0.00267 $ & $ \cdots $ & $ r$ &         HS\\
$ 55325.72598 $ & $  -0.00006 $ & $   0.00260 $ & $ \cdots $ & $ r$ &         HS\\
$ 55384.31621 $ & $  -0.00972 $ & $   0.00255 $ & $ \cdots $ & $ r$ &         HS\\
$ 55294.70845 $ & $   0.00056 $ & $   0.00244 $ & $ \cdots $ & $ r$ &         HS\\
$ 55263.69047 $ & $   0.00764 $ & $   0.00313 $ & $ \cdots $ & $ r$ &         HS\\
$ 55239.56544 $ & $   0.00769 $ & $   0.00264 $ & $ \cdots $ & $ r$ &         HS\\
$ 55301.60184 $ & $   0.00695 $ & $   0.00246 $ & $ \cdots $ & $ r$ &         HS\\
$ 55277.47684 $ & $   0.00689 $ & $   0.00280 $ & $ \cdots $ & $ r$ &         HS\\
$ 55287.81767 $ & $   0.00245 $ & $   0.00299 $ & $ \cdots $ & $ r$ &         HS\\

				[-1.5ex]
\enddata \tablenotetext{a}{
		The out-of-transit level has been subtracted. For the HATSouth light
		curve (rows with ``HS'' in the Instrument column), these magnitudes
		have been detrended using the EPD and TFA procedures prior to fitting a
		transit model to the light curve. Primarily as a result of this
		detrending, but also due to blending from neighbors, the apparent
		HATSouth transit depth is $\sim 79\%$ that of the true depth in the
		Sloan~$r$ filter. For the follow-up light curves (rows with an
		Instrument other than ``HS'') these magnitudes have been detrended with
		the EPD and TFA procedures, carried out simultaneously with the transit
		fit (the transit shape is preserved in this process).
} \tablenotetext{b}{
	Raw magnitude values without application of the EPD and TFA procedures.
	This is only reported for the follow-up light curves.
} \tablecomments{
	This table is available in a machine-readable form in the online journal.
	A portion is shown here for guidance regarding its form and content.
} \ifthenelse{\boolean{emulateapj}}{ \end{deluxetable*} }{ \end{deluxetable} }

\section{Analysis}
\label{sec:analysis}

\subsection{Properties of the parent star}
\label{sec:stelparam}

As mentioned in \refsecl{fiesobs} we determine the atmospheric parameters of
the host star \hatcur{}, including the effective temperature $\teffstar$, the
metallicity $\feh$, the projected rotational velocity $\vsini$, and the
stellar surface gravity $\loggstar$, by applying SPC to the FIES observations
of this star. This analysis yielded the following {\em initial} values and
uncertainties: $\teffstar=\hatcurSPCiteff$\,K, $\feh=\hatcurSPCizfeh$\,dex,
$\vsini=\hatcurSPCivsin\,\kms$, and $\loggstar=\hatcurSPCilogg$\,(cgs).

The same parameters were independently determined by applying SME to the
FEROS observations. The values obtained were $\teffstar=5790\pm120$\,K,
$\feh=-0.11\pm0.1$\,dex, $\vsini=4.8\pm0.5,\kms$, and
$\loggstar=4.53\pm0.17$\,(cgs). These measurements, apart from $\vsini$ are
consistent within the quoted error bars. For the remainder of the paper the
SPC results are adopted. The apparent inconsistency in the projected stellar
rotation velocity is likely due to the mis--estimation of the quoted errors.
For such slowly rotating stars, the uncertainties in both measurements are
significantly larger than $0.5\,\kms$.

Following \citet{sozzetti:2007} we use the stellar mean density \rhostar,
together with \teffstar\ and \feh\ from SPC to constrain the fundamental
properties of the star (mass, radius, age and luminosity) based on the
Yonsei-Yale \citep[Y2;][]{yi:2001} stellar evolution models. The stellar mean
density is closely related to the normalized semimajor axis \arstar, and is
determined by modeling the light curve (\refsec{globmod}). The light curve
model also depends on the quadratic limb darkening coefficients, which we
initially fix to the values from the \citet{claret:2004} law using the
initial \teffstar, \feh, and \loggstar\ values. The comparison with the
stellar evolution models provides a more precise estimate of $\loggstar$ than
can be determined from the spectrum. We we then fix $\loggstar$ to this more
precise value in a second iteration of SPC to determine revised estimates of
the other atmospheric parameters ($\teffstar$, $\feh$, and $\vsini$). These
new atmospheric parameters are used to determine new estimates of the limb
darkening coefficients, which are used in a second iteration of the light
curve modeling. We find that the value of $\loggstar$ converges after two
iterations.

\begin{figure}[!ht]
\plotone{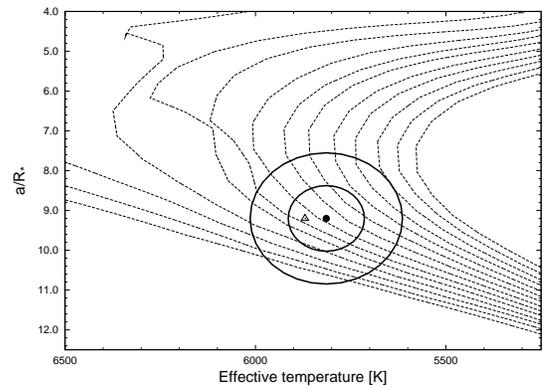}
\caption{
    Model isochrones from \cite{\hatcurisocite} for the measured
    metallicity of \hatcur, \feh = \hatcurSPCiizfehshort, and ages of
    0.2\,Gyr, and 1 to 13\,Gyr in 1\,Gyr increments (left to right).
    The adopted values of $\teffstar$ and \arstar\ are shown together
    with their 1$\sigma$ and 2$\sigma$ confidence ellipsoids. The
    open triangle shows the values from the initial SPC iteration.
\label{fig:iso}}
\end{figure}

We collect the final values for the observed and derived stellar parameters
for \hatcur{} in \reftabl{stellar}. The inferred location of the star in a
diagram of \arstar\ versus \teffstar, analogous to the classical H-R
diagram, is shown in \reffigl{iso}. The stellar properties and their
1$\sigma$ and 2$\sigma$ confidence ellipsoids are displayed against
the backdrop of \cite{\hatcurisocite} isochrones for the measured
metallicity of \feh\ = \hatcurSPCiizfehshort, and a range of ages.


\begin{deluxetable}{lrl}
\tablewidth{0pc}
\tabletypesize{\scriptsize}
\tablecaption{
	Stellar parameters for \hatcur{}
	\label{tab:stellar}
}
\tablehead{
	\colhead{~~~~~~~~Parameter~~~~~~~~}	&
	\colhead{Value} &
	\colhead{Source}
}
\startdata
\noalign{\vskip -3pt}
\sidehead{Spectroscopic properties}
~~~~$\teffstar$ (K)\dotfill         &  \hatcurSPCteff       & SPC\tablenotemark{a}\\
~~~~$\feh$\dotfill                  &  \hatcurSPCzfeh       & SPC                 \\
~~~~$\vsini$ (\kms)\dotfill         &  \hatcurSPCvsin       & SPC                 \\
\sidehead{Photometric properties}
~~~~$V$ (mag)\dotfill               &  \hatcurCCapassmV      & APASS                \\
~~~~$B$ (mag)\dotfill       &  \hatcurCCapassmB      & APASS                \\
~~~~$g$ (mag)\dotfill       &  \hatcurCCapassmg      & APASS                \\
~~~~$r$ (mag)\dotfill       &  \hatcurCCapassmr      & APASS                \\
~~~~$i$ (mag)\dotfill       &  \hatcurCCapassmi      & APASS                \\
~~~~$J$ (mag)\dotfill               &  \hatcurCCtwomassJmag & 2MASS           \\
~~~~$H$ (mag)\dotfill               &  \hatcurCCtwomassHmag & 2MASS           \\
~~~~$K_s$ (mag)\dotfill             &  \hatcurCCtwomassKmag & 2MASS           \\
\sidehead{Derived properties}
~~~~$\mstar$ ($\msun$)\dotfill      &  \hatcurISOmlong      & \hatcurisoshort+\hatcurlumind+SPC \tablenotemark{b}\\
~~~~$\rstar$ ($\rsun$)\dotfill      &  \hatcurISOrlong      & \hatcurisoshort+\hatcurlumind+SPC         \\
~~~~$\loggstar$ (cgs)\dotfill       &  \hatcurISOlogg       & \hatcurisoshort+\hatcurlumind+SPC         \\
~~~~$\lstar$ ($\lsun$)\dotfill      &  \hatcurISOlum        & \hatcurisoshort+\hatcurlumind+SPC         \\
~~~~$M_V$ (mag)\dotfill             &  \hatcurISOmv         & \hatcurisoshort+\hatcurlumind+SPC         \\
~~~~$M_K$ (mag,\hatcurjhkfilset)\dotfill &  \hatcurISOMK    & \hatcurisoshort+\hatcurlumind+SPC         \\
~~~~Age (Gyr)\dotfill               &  \hatcurISOage        & \hatcurisoshort+\hatcurlumind+SPC         \\
~~~~Distance (pc)\dotfill           &  \hatcurXdist         & \hatcurisoshort+\hatcurlumind+SPC\\
[-1.5ex]
\enddata
\tablenotetext{a}{
	SPC = ``Spectroscopic Parameter Classification'' procedure
        applied to the high-resolution FIES spectra (Buchhave et al.,
        in preparation).  These parameters rely primarily on SPC, but
        have a small dependence also on the iterative analysis
        incorporating the isochrone search and global modeling of the
        data, as described in the text.  }
\tablenotetext{b}{
	\hatcurisoshort+\hatcurlumind+SPC = Based on the \hatcurisoshort\
    isochrones \citep{\hatcurisocite}, \hatcurlumind\ as a luminosity
    indicator, and the SPC results.
}
\end{deluxetable}

\subsection{Excluding blend scenarios}
\label{sec:blend}

Most of the obvious astrophysical false positive scenarios which might
produce the detected transit of \hatcur{} are ruled out by the spectroscopic
observations discussed in \refsecl{spec}. Some scenarios involving diluted
eclipsing binary systems, or transiting planet systems diluted by an
additional star, may still be consistent with these observations. To rule
these scenarios out we conduct detailed modeling of the light curves
following the procedure described in \citet{hartman:2011}.

From the light curves alone we are able to rule out hierarchical
triple star systems with greater than $4\sigma$ confidence, and blends
between a foreground star and a background eclipsing binary with
$\gtrsim 2.5\sigma$ confidence. Moreover, the only nonplanetary blend
scenarios which could plausibly fit the light curves (scenarios for
which the rejection confidence is less than $5\sigma$) are scenarios
in which the two brightest stars in the blended system are nearly
equal in brightness; such scenarios would have been easily detected
from the spectroscopic observations (the spectrum would have either
been obviously double-lined, or there would have been several
\kms\ changes in the RV, BS, and/or FWHM; the latter is computed, and
checked for variations, but not shown).

While our analysis rules out nonplanetary explanations for the
observations, we are unable to disprove the possibility that the
object is an unresolved stellar binary system (or chance alignment of
two stars) with one component hosting a transiting planet. If this is
the case the true mass and radius of the planet would be larger than
what we infer. However, lacking any positive evidence for multiple
stellar components in the system, for the analysis conducted in
\refsecl{globmod} we assume that \hatcur{} consists of a single star
with a transiting planet.

\subsection{Global modeling of the data}
\label{sec:globmod}

We conducted a joint modeling of the HATSouth photometry, follow-up
photometry, and RV data following the procedure described by
\citet{bakos:2010:hat11}. Briefly, we modeled the light curves using
analytic formulae based on \citet{mandel:2002} to describe the
physical variation together with EPD and TFA to describe instrumental
variations, and we modeled the RV data with a Keplerian orbit with
the formalism of \citet{pal:2009:orbits}. We used the Downhill Simplex
Algorithm \citep[e.g.][]{press:1992} to optimize the parameters,
followed by a Markov-Chain Monte Carlo analysis \citep[MCMC,
  e.g.][]{ford:2006} to determine their uncertainties and
correlations. The resulting geometric parameters pertaining to the light
curves and velocity curves, as well as the derived physical planetary
parameters, are listed in \reftabl{planetparam}.

Included in this table is the RV ``jitter'' which was added in
quadrature to the formal RV errors for the FEROS observations, such
that $\chi^{2}/{\rm dof} = 1$ for this instrument. The origin of this
noise is unknown, it may either be to due astrophysical noise
intrinsic to the star (potentially including additional planets in the
system), or instrumental errors not included in the formal
uncertainties.

We find that the planet has a mass of $\mpl=\hatcurPPmlong\,\mjup$ and
radius of $\rpl=\hatcurPPrlong\,\rjup$, which are typical values for a
hot Jupiter. We allow the eccentricity to vary in the fit so as to
allow our uncertainty on this parameter to be included in the
uncertainties of the other physical parameters of the system, however
we find $e = \hatcurRVeccen$, which is consistent with zero
eccentricity. Our $99\%$ upper-limit on the eccentricity is $e <
0.4$.

\begin{deluxetable}{lr}
\tabletypesize{\scriptsize}
\tablecaption{Orbital and planetary parameters\label{tab:planetparam}}
\tablehead{
	\colhead{~~~~~~~~~~~~~~~Parameter~~~~~~~~~~~~~~~} &
	\colhead{Value}
}
\startdata
\noalign{\vskip -3pt}
\sidehead{\Lc{} parameters}
~~~$P$ (days)             \dotfill    & $\hatcurLCP$              \\
~~~$T_c$ (${\rm BJD}$)    
      \tablenotemark{a}   \dotfill    & $\hatcurLCT$              \\
~~~$T_{14}$ (days)
      \tablenotemark{a}   \dotfill    & $\hatcurLCdur$            \\
~~~$T_{12} = T_{34}$ (days)
      \tablenotemark{a}   \dotfill    & $\hatcurLCingdur$         \\
~~~$\arstar$              \dotfill    & $\hatcurPPar$             \\
~~~$\zrstar$\tablenotemark{b}              \dotfill    & $\hatcurLCzeta$           \\
~~~$\rpl/\rstar$          \dotfill    & $\hatcurLCrprstar$        \\
~~~$b \equiv a \cos i/\rstar$
                          \dotfill    & $\hatcurLCimp$            \\
~~~$i$ (deg)              \dotfill    & $\hatcurPPi$              \\

\sidehead{Limb-darkening coefficients \tablenotemark{c}}
~~~$a_g$ (linear term)   \dotfill    & $\hatcurLBig$             \\
~~~$b_g$ (quadratic term) \dotfill    & $\hatcurLBiig$            \\
~~~$a_r$                 \dotfill    & $\hatcurLBir$             \\
~~~$b_r$                 \dotfill    & $\hatcurLBiir$            \\
~~~$a_i$                 \dotfill    & $\hatcurLBii$             \\
~~~$b_i$                  \dotfill    & $\hatcurLBiii$            \\

\sidehead{RV parameters}
~~~$K$ (\ms)              \dotfill    & $\hatcurRVK$              \\
~~~$e\cos\omega$ 
                          \dotfill    & $\hatcurRVk$              \\
~~~$e\sin\omega$
                          \dotfill    & $\hatcurRVh$              \\
~~~$e$                    \dotfill    & $\hatcurRVeccen$          \\
~~~RV jitter (\ms)\tablenotemark{d}        
                          \dotfill    & $\hatcurRVjitterB$          \\

\sidehead{Planetary parameters}
~~~$\mpl$ ($\mjup$)       \dotfill    & $\hatcurPPmlong$          \\
~~~$\rpl$ ($\rjup$)       \dotfill    & $\hatcurPPrlong$          \\
~~~$C(\mpl,\rpl)$
    \tablenotemark{e}     \dotfill    & $\hatcurPPmrcorr$         \\
~~~$\rhopl$ (\gcmc)       \dotfill    & $\hatcurPPrho$            \\
~~~$\log g_p$ (cgs)       \dotfill    & $\hatcurPPlogg$           \\
~~~$a$ (AU)               \dotfill    & $\hatcurPParel$           \\
~~~$T_{\rm eq}$ (K)       \dotfill    & $\hatcurPPteff$           \\
~~~$\Theta$\tablenotemark{f}\dotfill  & $\hatcurPPtheta$          \\
~~~$\langle F \rangle$ ($10^{\hatcurPPfluxavgdim}$\ergscmsq) 
\tablenotemark{g}         \dotfill    & $\hatcurPPfluxavg$        \\
[-1.5ex]
\enddata
\tablenotetext{a}{
    \ensuremath{T_c}: Reference epoch of mid transit that minimizes the
    correlation with the orbital period. BJD is calculated from UTC.
	\ensuremath{T_{14}}: total transit duration, time between first to
	last contact;
	\ensuremath{T_{12}=T_{34}}: ingress/egress time, time between first
	and second, or third and fourth contact.
}
\tablenotetext{b}{
    Reciprocal of the half duration of the transit used as a jump
    parameter in our MCMC analysis in place of $\arstar$. It is
    related to $\arstar$ by the expression $\zrstar = \arstar
    (2\pi(1+e\sin \omega))/(P \sqrt{1 - b^{2}}\sqrt{1-e^{2}})$
    \citep{bakos:2010:hat11}.
}
\tablenotetext{c}{
	Values for a quadratic law given separately for the Sloan~$g$,
        $r$, and $i$ filters. These values were adopted from the
        tabulations by \cite{claret:2004} according to the
        spectroscopic (SPC) parameters listed in \reftabl{stellar}.
}
\tablenotetext{d}{
    This jitter was added to the FEROS measurements only. Formal
    uncertainties for the Coralie measurements were not determined
    from the pipeline, instead the uncertainties were fixed such that
    $\chi^{2}/{\rm dof} = 1$ for these observations, so that, by
    definition, no jitter was required.
}
\tablenotetext{e}{
	Correlation coefficient between the planetary mass \mpl\ and radius
	\rpl.
}
\tablenotetext{f}{
	The Safronov number is given by $\Theta = \frac{1}{2}(V_{\rm
	esc}/V_{\rm orb})^2 = (a/\rpl)(\mpl / \mstar )$
	\citep[see][]{hansen:2007}.
}
\tablenotetext{g}{
	Incoming flux per unit surface area, averaged over the orbit.
}
\end{deluxetable}



\section{Discussion}
\label{sec:discussion}

The HATSouth global network of telescopes relies on combining observations
from three stations located at three different sites in the southern
hemisphere, with a close to optimal longitude separations. This makes it
possible to continuously (or nearly so if the night is short) observe a given
portion of the sky to search for transits of planets in front of their parent
stars.

In this paper we have presented the first planet discovered by the HATSouth
network. This discovery demonstrates that we are indeed able to successfully
combine observations from multiple telescopes and sites to detect planetary
transit signals.

In many respects this is a very typical transiting planet, with an orbital
period of $P\approx\hatcurLCPshort$\,days (close to the mode for transiting
planets detected from the ground), mass $\mpl \approx
\hatcurPPmshort$\,\mjup\ (slightly higher than typical), radius $\rpl \approx
\hatcurPPrshort$\,\rjup\ (close to the median for its mass), around a star
that is an almost exact analogue of the Sun. 

\begin{figure*}[!ht]
\plottwo{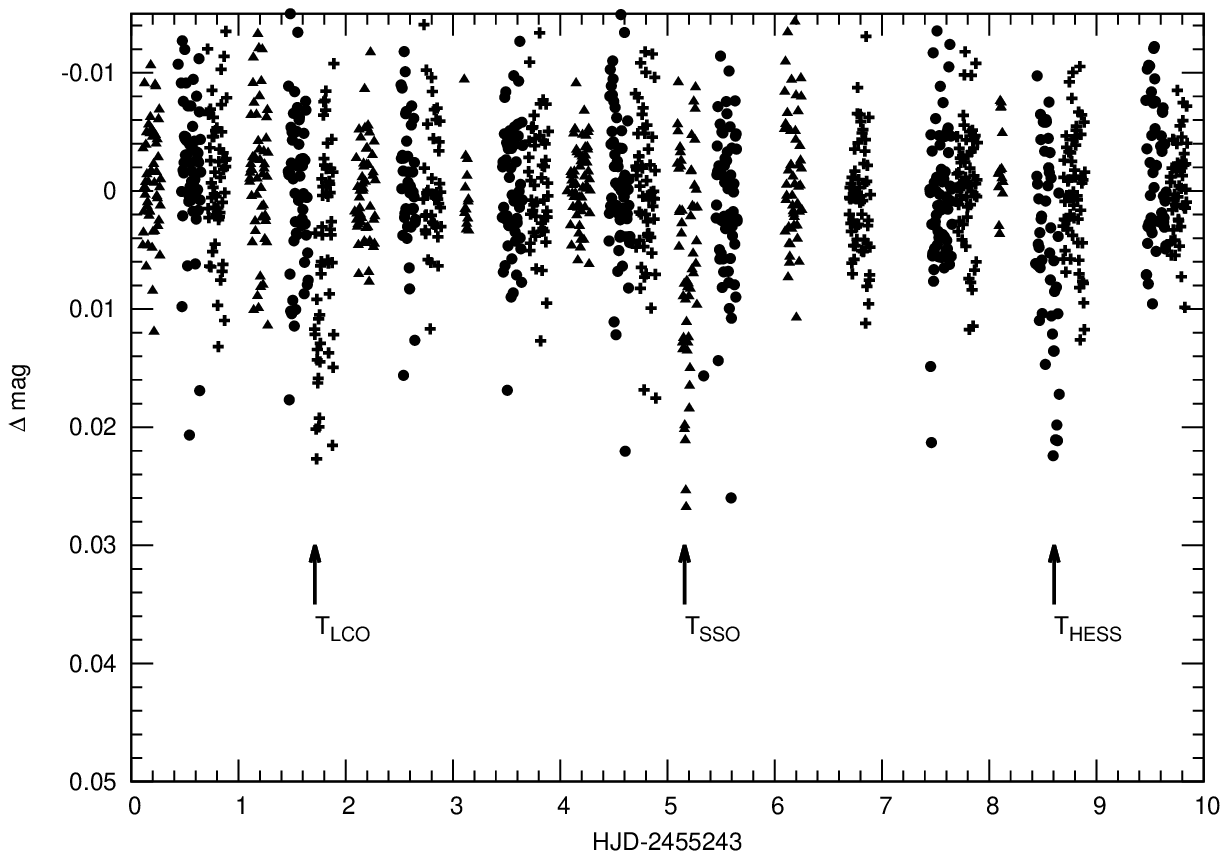}{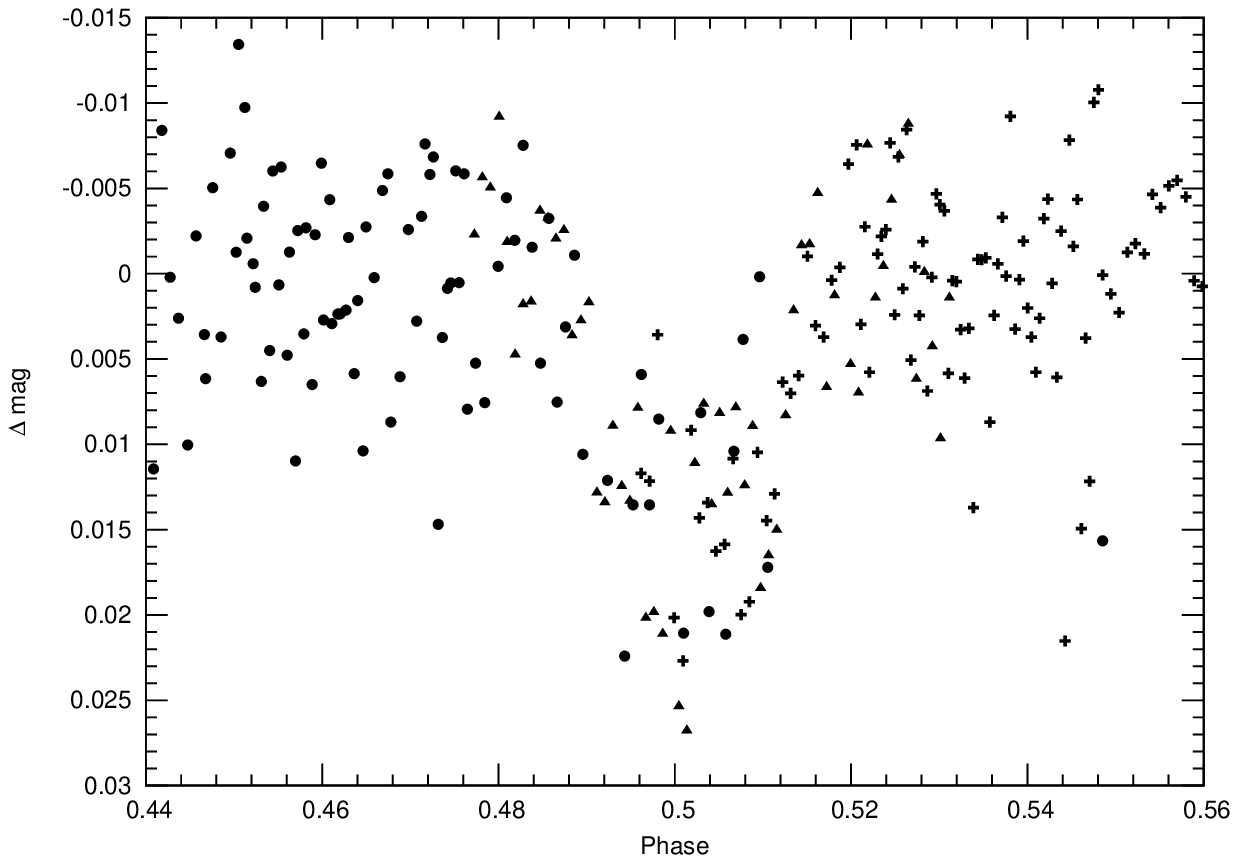}
\caption{
    (Left): The deviations of the measured brightness of \hatcur\ from the
    median measured in magnitudes as a function of time in days. The
    different symbols denote observations from different sites/stations:
    pluses from LCO, circles from HESS and triangles from SSO.
    (Right): The same plot, but this time phase folded and zoomed in around the
    transit, occurring at phase of 0.5.
\label{fig: HS 10days}}
\end{figure*}

Figure \ref{fig: HS 10days} demonstrates the power of using a network of
telescopes to search for transit signals among stars. It shows a ten day
period (starting 2010 February 15) from the HATSouth discovery light curve
containing 1351 observations of \hatcur, i.e. a small fraction of the total
of more than 12\,000 observations. Over this period, a significant
fraction of the observations come from each of the three HATSouth sites, and
a transit is detected by each station at a comparable significance. Despite
the short nights in the southern hemisphere at this time, near continuous
coverage is achieved and every single transit in this time window is observed
by the HATSouth network. In contrast, a search operating from a single
location would only be able to detect one of these transits. 

As discussed in \citet{bakos:2012:hs} this increased duty cycle should result
in many more planet detections, especially at longer periods and for smaller
planetary radii, which are arguably the most valuable. 

In \citet{bakos:2012:hs} detailed simulations were carried out showing that
the distribution of planets expected to be produced by the HATSouth survey
peaks at planet radii between 1 and 2 Jupiter radii and orbits with periods
between 3 and 5 days. Not surprisingly, the first planetary system discovered
by the network falls within these ranges. However, as discussed in that
paper, these predicted distributions predict significantly enhanced
sensitivity to planets with much longer orbital periods and smaller radii
compared to a single site survey. 

\acknowledgements 

Development of the HATSouth project was funded by NSF MRI grant
NSF/AST-0723074, operations are supported by
NASA grant NNX09AB29G, and follow-up observations receive partial support
from grant NSF/AST-1108686.
Followup observations with the ESO~2.2\,m/FEROS instrument were performed
under MPI guaranteed time (P087.A-9014(A), P088.A-9008(A), P089.A-9008(A))
and Chilean time (P087.C-0508(A)).
A.J.\ acknowledges support from Fondecyt project 1095213, Ministry of Economy
ICM Nuclei P07-021-F and P10-022-F, Anillo ACT-086 and BASAL CATA PFB-06.
V.S.\ acknowledges support form BASAL CATA PFB-06.  M.R.\ acknowledges
support from a FONDECYT postdoctoral fellowship. R.B.\ and N.E.\ acknowledge
support from Fondecyt project 1095213.
This work is 
based on observations made with
ESO Telescopes at the La Silla Observatory under programme IDs P087.A-9014(A),
P088.A-9008(A), P089.A-9008(A), P087.C-0508(A), 089.A-9006(A), and
observations made with the
Nordic Optical Telescope, operated on the island of La Palma jointly
by Denmark, Finland, Iceland, Norway, and Sweden, in the Spanish
Observatorio del Roque de los Muchachos of the Instituto de
Astrofisica de Canarias. 
This
paper also uses observations obtained with facilities of the Las Cumbres
Observatory Global Telescope.
Work at the Australian National University is supported by ARC Laureate 
Fellowship Grant FL0992131.
We acknowledge the use of the AAVSO Photometric All-Sky
Survey (APASS), funded by the Robert Martin Ayers Sciences Fund, and 
the SIMBAD database,
operated at CDS, Strasbourg, France. 
M.~Mohler wants to thank N.~Piskunov and his group in Uppsala for
introducing her to Spectroscopy Made Easy (SME).
Operations at the MPG/ESO 2.2\,m Telescope are jointly performed by the
Max Planck Gesellschaft and the European Southern Observatory. The
imaging system GROND has been built by the high-energy group of MPE in
collaboration with the LSW Tautenburg and ESO\@. We thank Timo Anguita
and R\'egis Lachaume for their technical assistance during the
observations at the MPG/ESO 2.2\,m Telescope.
We thank Miguel Roth, Francesco Di Mille and Rodolfo Angeloni for allowing us
to use the Swope telescope on May 24/25 2011.
We are grateful to P.Sackett for her help in the early phase of
the HATSouth project.


\bibliographystyle{apj}
\bibliography{hatsbib}

\end{document}